\RequirePackage{fixltx2e}
\documentclass[article,A4,twocolumn,amsmath,superscriptaddress,showpacs,floatfix]{revtex4-1}

\usepackage{siunitx}
\usepackage{graphicx}
\usepackage{color}
\usepackage{epstopdf}
\usepackage{soul}

\begin{document}

%\title{Nonlinear interaction between single photons mediated by a single quantum dot in a photonic-crystal waveguide}
\title{Single-photon nonlinear optics with a quantum dot in a waveguide}

\author{A.~Javadi}
%\email{Javadi@nbi.ku.dk}
\affiliation{Niels Bohr Institute, University of Copenhagen, Blegdamsvej 17, DK-2100 Copenhagen, Denmark}
\author{I.~S\"{o}llner}
\affiliation{Niels Bohr Institute, University of Copenhagen, Blegdamsvej 17, DK-2100 Copenhagen, Denmark}
\author{M.~Arcari}
\affiliation{Niels Bohr Institute, University of Copenhagen, Blegdamsvej 17, DK-2100 Copenhagen, Denmark}
\author{S.~L.~Hansen}
\affiliation{Niels Bohr Institute, University of Copenhagen, Blegdamsvej 17, DK-2100 Copenhagen, Denmark}
\author{L. ~Midolo}
\affiliation{Niels Bohr Institute, University of Copenhagen, Blegdamsvej 17, DK-2100 Copenhagen, Denmark}
\author{S.~Mahmoodian}
\affiliation{Niels Bohr Institute, University of Copenhagen, Blegdamsvej 17, DK-2100 Copenhagen, Denmark}
\author{G. ~Kir{\v{s}}ansk{\.{e}}}
\affiliation{Niels Bohr Institute, University of Copenhagen, Blegdamsvej 17, DK-2100 Copenhagen, Denmark}
\author{T. ~Pregnolato}
\affiliation{Niels Bohr Institute, University of Copenhagen, Blegdamsvej 17, DK-2100 Copenhagen, Denmark}
\author{E.~H.~Lee}
\affiliation{Center for Opto-Electronic Convergence Systems, Korea Institute of Science and Technology, Seoul, 136-791, Korea}
\author{J.~D.~Song}
\affiliation{Center for Opto-Electronic Convergence Systems, Korea Institute of Science and Technology, Seoul, 136-791, Korea}
\author{S.~Stobbe}
\affiliation{Niels Bohr Institute, University of Copenhagen, Blegdamsvej 17, DK-2100 Copenhagen, Denmark}
\author{P.~Lodahl}\email{Lodahl@nbi.ku.dk} \homepage{http://quantum-photonics.nbi.ku.dk}
\affiliation{Niels Bohr Institute, University of Copenhagen, Blegdamsvej 17, DK-2100 Copenhagen, Denmark}

\date{\today}

\maketitle

{\bf Strong nonlinear interactions between photons enable logic operations for both classical and quantum-information technology. Unfortunately, nonlinear interactions are usually feeble and therefore all-optical logic gates tend to be inefficient. A quantum emitter deterministically coupled to a propagating mode fundamentally changes the situation, since each photon inevitably interacts with the emitter, and highly correlated many-photon states may be created {\cite{Rice1988IEEE,kojima2003PRA,shen2007strongly,Chang2014NPHOT,Lodahl2013arXivRMP}}. Here we show that a single quantum dot in a photonic-crystal waveguide can be utilized as a giant nonlinearity sensitive at the single-photon level. The nonlinear response is revealed from the intensity and quantum statistics of the scattered photons, and contains contributions from an entangled photon-photon bound state. The quantum nonlinearity will find immediate applications for deterministic Bell-state measurements {\cite{Witthaut2012EL}} and single-photon transistors \cite{Chang2007NPHYS} and paves the way to scalable waveguide-based photonic quantum-computing architectures  {\cite{Zheng2013PRL,Duan2004PRL}}\rm.

Optical photons are excellent carriers of information over extended distances since they can be distributed fast and efficiently. The access to nonlinearity enables the processing of information stored in light. An efficient nonlinearity capable of operating down to the ultimate level of single photons has been long sought after, as it would open new avenues for optical signal processing \cite{GibbsBook2012} and could improve linear-optics quantum-information architectures \cite{Knill2001Nature}. Photonic nanostructures provide a route to overcoming these limitations since light and matter can be deterministically interfaced. One approach to photon nonlinearities exploits the anharmonic spectrum of a cavity polariton, which has been experimentally demonstrated both with atoms {\cite{birnbaum2005NPHOT}} and quantum dots {\cite{Srinavasan2007Nature,Faraon2008NPHYS,Reinhard2012NPHOT,Loo2012PRL,Kim2013NPHOT}}. An alternative approach uses the intrinsic nonlinearity of a quantum emitter deterministically coupled to a single photonic mode (a `1D atom'); such a coupling was recently achieved with single quantum dots in photonic-crystal waveguides \cite{Arcari2014PRL}. Theoretical studies have predicted that intricate photon-photon \cite{shen2007strongly} and photon-emitter  \cite{Longo2010PRL} entanglement (so-called bound states) may be induced by deterministic few-photon scattering. So far a quantum-emitter nonlinearity has been observed at microwave frequencies with superconducting qubits \cite{Hoi2012PRL} and at optical frequencies with single atoms in cavities \cite{Tiecke2014Nature,volz2014nonlinear}. The present work reports on single-photon interference and nonlinear scattering of optical photons in a scalable solid-state platform.

A quantum dot in a photonic waveguide is a particularly attractive approach to quantum nonlinear optics since it can be naturally incorporated in integrated photonic circuits.
In the setting of cavity QED, scaling to larger coupled quantum systems is demanding since each individual emitter and cavity must be precisely tuned. The very wide coupling bandwidth in photonic-crystal waveguides \cite{Lund-Hansen2008PRL} implies that scalable quantum architectures may be envisioned with significantly less experimental overhead, since only the quantum dot needs to be tuned.   %Furthermore, the requirement of active tuning of the  quantum-dot energy levels is much less stringent than in a cavity due to the very wide bandwidth (more than 20 $\mathrm{nm}$) of the slow-light regime in a photonic-crystal waveguide  \cite{Lund-Hansen2008PRL}.
A quantum dot in a photonic-crystal waveguide constitutes the paradigmatic example of a 1D artificial atom where light-matter interaction is fundamentally different from that in 3D. For instance, dipole-induced single-photon interference may be studied, which can be considered a precursor to the experimental demonstration of single-photon nonlinearity and photon-photon bound states. Furthermore, an anomalous radiative Lamb shifts or infinitely ranging dipole-dipole interaction have been been predicted \cite{Lodahl2013arXivRMP}.

\begin{figure*}[t!]
\includegraphics[width=180mm]{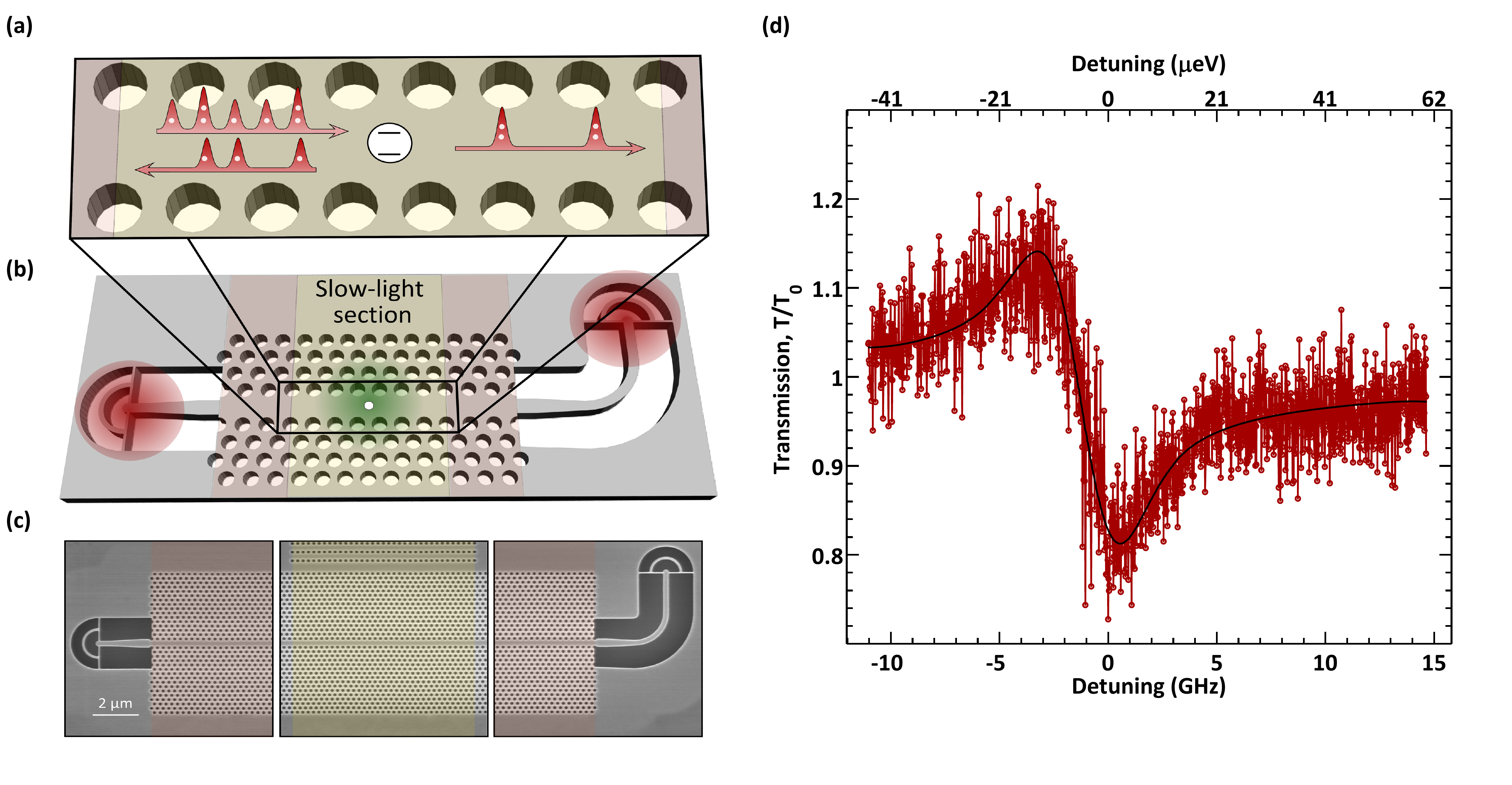}
\caption{\label{fig_1}|\textbf{Resonant spectroscopy on a quantum dot in a photonic-crystal waveguide.} (a) Operational principle of resonant scattering in a photonic-crystal waveguide. Single-photon components of the incoming light are reflected by the quantum dot while two- and more-photon components are preferentially transmitted. (b) Illustration and (c) scanning-electron micrograph of the sample. A quantum dot (white circle) in the central part of the slow-light section is excited by launching light through one grating and detecting light from the other grating. The red areas indicate the size of the excitation and collection areas. The green area is the illumination region of the repump laser, see SI. (d) Resonant transmission spectrum recorded by scanning a narrow-band continuous-wave laser through the resonance of a quantum dot in the case of weak excitation. The power on the sample was \SI{50}{pW}, which is far below the critical power of \SI{1.9}{nW}. The solid black line is a model of the experimental data of the Fano resonance, see SI for details of the model. }
\end{figure*}

The operational principle of the quantum-dot nonlinearity is outlined in Fig.\ \ref{fig_1}(a): by scattering a weak resonant laser on the quantum dot in the photonic-crystal waveguide, the single-photon component is reflected while two- and higher-photon components have an increased probability of being transmitted. The layout and scanning electron micrograph of the sample are shown in Figs. \ref{fig_1}(b) and (c), respectively. The sample consists of a central slow-light waveguide section (slow-down factor {$n_g \sim 30$}) terminated on each side by two waveguide sections ({$n_g = 5$}) and coupled through suspended waveguides to two gratings, which direct the emission vertically out of the planar structure. One grating is used for launching light into the waveguide and the other for extracting the transmitted light. Furthermore, the quantum dot is exposed to a repump laser in order to prepare and stabilize the initial state of the emitter before the scattering process. The detailed description of the sample and experimental procedure is presented in the Supplementary Information (SI). Figure \ref{fig_1}(d) shows an example of the transmission spectrum recorded when scanning a narrow-linewidth laser through the resonance feature of a quantum dot coupled to a photonic-crystal waveguide. These measurements are performed at low excitation powers in the coherent-scattering regime where the incoming and outgoing fields maintain a fixed phase relation. Residual reflections from the waveguide ends imply that weak Fabry-Perot resonances form, which lead to a characteristic Fano line shape \cite{Shen2005OL, Goban2014NCOM} of $\sim 30\%$ modulation. The peak (dip) of the transmission spectrum corresponds to dipole-induced transparency (reflection) resulting from single-photon interference. The size of the modulation is determined by the  $\beta$-factor, which quantifies the coupling efficiency, as well as the spectral diffusion and blinking of the quantum-dot line due to charge or spin noise {\cite{Kuhlmann2013NPHYS}}, and residual broadening of the zero-phonon line, e.g., due to carrier-phonon interactions \cite{Muljarov2004PRL}. Spectral diffusion and blinking processes can be strongly reduced by implementing electrical gates on the quantum-dot samples {\cite{Kuhlmann2013NPHYS}} and can be further suppressed through active stabilization.

\begin{figure}[t!]
\includegraphics[width=\columnwidth]{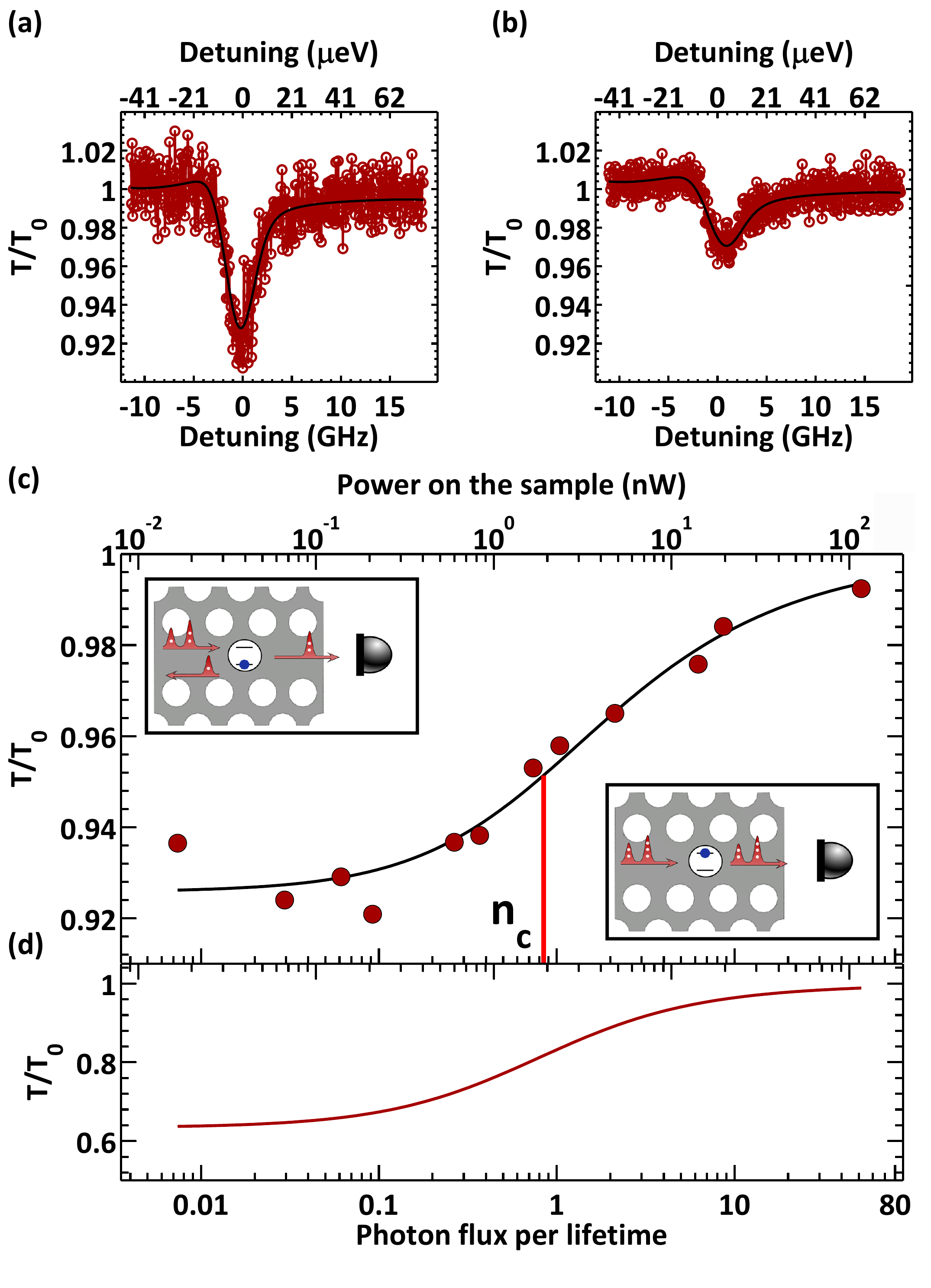}
\caption{\label{trans_fig}| \textbf{Nonlinear response of a single quantum dot in a photonic-crystal waveguide}. Examples of transmission spectra recorded at two different powers of (a) {$P=0.18 \: \mathrm{nW}$} and (b) {$P=2.2 \: \mathrm{nW}$}. (c) Transmission on resonance with the quantum dot versus incident photon flux relative to the emitter lifetime. The top axis shows the corresponding optical power applied to the sample. The solid line is a fit to the experimental data. The critical power that characterizes the saturation curve is indicated on the axis. The insets show the measurement geometry and illustrate that for weak excitation the quantum dot preferentially reflects while it becomes transparent at stronger excitation where two- and higher-photon components of the coherent state dominate.  (d) Same as (c) but after deconvolution of the spectral diffusion and blinking.}
\end{figure}

The nonlinear response of the quantum dot is investigated by recording the transmission as a function of excitation power. Two examples of transmission spectra are displayed in Figs. \ref{trans_fig}(a) and \ref{trans_fig}(b) for weak and intermediate power, respectively. In this data set a transmission dip of $\sim8 \%$ was recorded, limited by spectral diffusion, which was found to vary when the sample was heated up and subsequently cooled down.  Figure \ref{trans_fig}(c) shows the  transmission as a function of power inside the waveguide and displays a characteristic nonlinear saturation behavior.  The data are modelled very well by the theory of Ref. \cite{Auffeves-Garnier2007PRA} (cf. SI for a detailed account) for a coupling efficiency of $\beta=85\%$, broadening by spectral diffusion of $\sigma/\Gamma=3.6$, blinking probability of $\alpha=0.43$, and a pure dephasing rate describing the broadening of the zero-phonon line of $\gamma_0/\Gamma=0.79$. Here the emitter decay rate  $\Gamma = \SI{2.5}{\nano\second}^{-1}$ is obtained in time-resolved measurements leading to an independent measurement of the coupling efficiency of $\beta \sim 96 \%$ \cite{Arcari2014PRL}. The two values are consistent since the $\beta$-factor extracted from resonant scattering experiments is effectively reduced compared to the value obtained from the dynamics by the presence of weak phonon sidebands due to incoherent Raman scattering processes \cite{Konthasinghe2012PRB}. Figure \ref{trans_fig}(d) shows the nonlinear response after deconvolution of the spectral diffusion and blinking, which can be eliminated by implementing electrical gates. We extract a critical photon flux per lifetime of {$n_c = 0.81$} characterizing the nonlinear saturation curve, demonstrating that the nonlinearity operates at the ultimate level of single photons. This corresponds to a characteristic switching energy of only $\sim 0.17$ attojoule. For comparison, the actual power applied to the sample at the critical power is {\SI{1.9}{nW}}, which implies that {23 \%} of the excitation beam is successfully coupled to the quantum dot in the waveguide. Photonic waveguide nonlinearities are particularly promising for obtaining nonlinearities at low photon numbers; a detailed comparison to the case of cavity polaritons is presented in SI.

 \begin{figure}[ht!]
\includegraphics[width=\columnwidth]{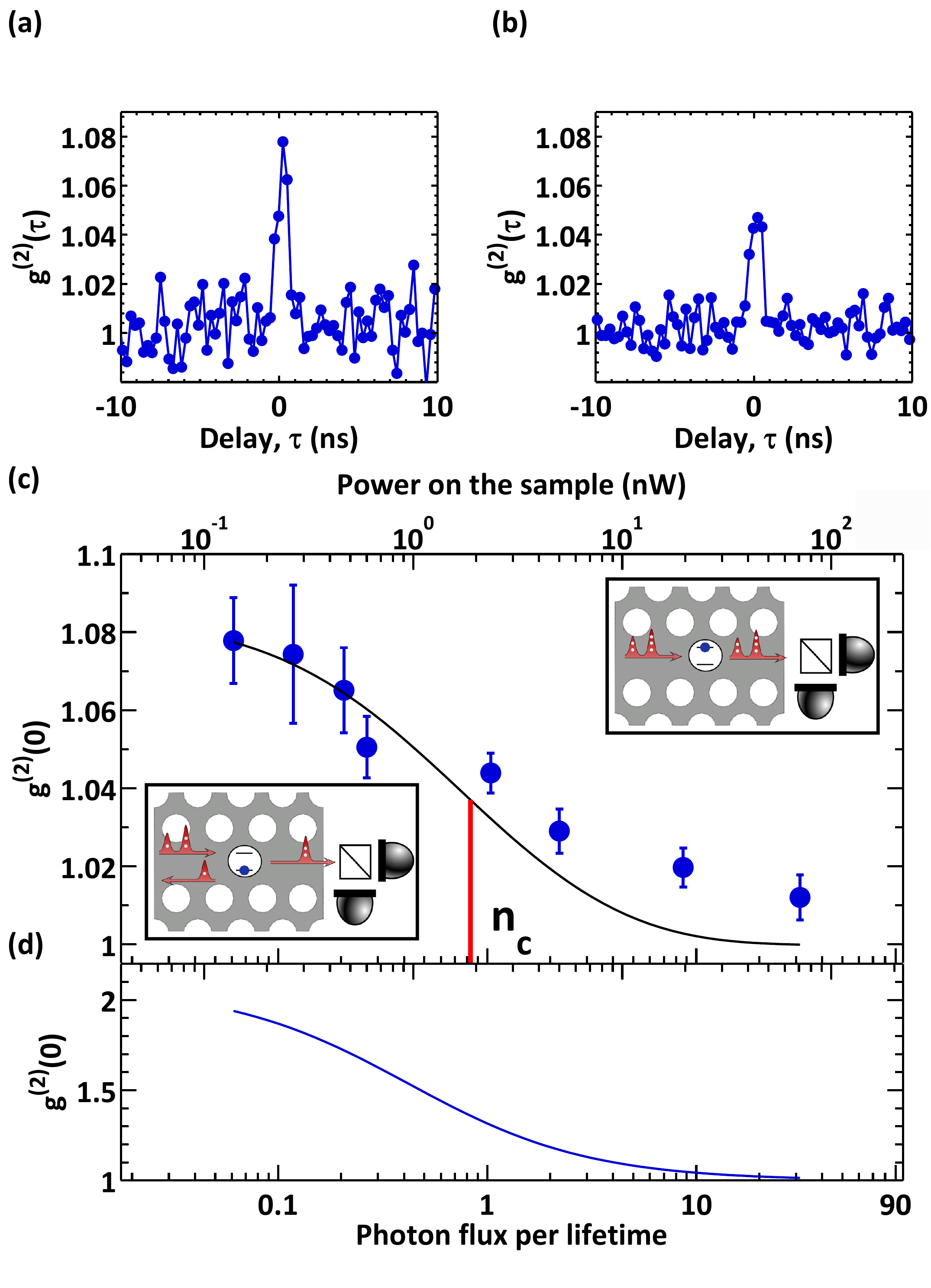}
\caption{\label{g2_fig}| \textbf{Photon statistics induced by the single-photon nonlinearity}. (a) and (b) Measurements of the autocorrelation function of the transmitted light recorded on resonance with the quantum dot for the conditions corresponding to Fig. \ref{trans_fig} (a) and (b), respectively. (c) Power dependence of the autocorrelation function peak at $\tau=0$. The solid black line is a fit to the data. A maximum bunching of $8 \%$ is observed, which corresponds to $18 \%$ when accounting for the finite response time of the detection. The vertical line indicates the critical power. (d) Same as (c) but after deconvolution of the spectral diffusion and blinking.  }
\end{figure}

By monitoring the photon statistics of the transmitted light, the single-photon character of the nonlinear response and the ability to generate photon-photon bound states are revealed. In the ideal case of $\beta \rightarrow 1$ and for a weak coherent state, the single-photon component is fully reflected while two- and higher-photon components are preferentially transmitted leading to photon bunching in the transmission \cite{Rice1988IEEE,Chang2007NPHYS}. Figure \ref{g2_fig}(a) and \ref{g2_fig}(b) show the experimental signature of photon bunching in the intensity autocorrelation function $g^2(\tau)$ of the transmission as a function of the delay $\tau$. The peak centered at $\tau =0 $ is the experimental signature that two or more photons impinging on the quantum dot within its radiative lifetime interact leading to photon-photon bound states. We estimate a photon-photon bound-state contribution to the transmission probability of the two-photon component of $\sim 70 \%$, which is the entangled part of the scattered light (see SI for further details). The observed photon bunching is highly sensitive to decoherence processes \cite{Lodahl2013arXivRMP} and the experimental observation of this effect testifies that resonant photon scattering using quantum dots is highly coherent \cite{Matthiesen2013NCOM}.  The amount of photon bunching is found to decrease with increasing power (cf. Fig. \ref{g2_fig}(c)) since two- and higher-photon components increasingly dominate the input state. At high excitation power, we observe $g^2(0) \approx 1$, corresponding to the Poissonian statistics of a coherent state, i.e., the recorded light is unaffected by the saturated quantum dot. This experiment demonstrates the basic operational principle behind photon sorting with  applications in photonic quantum-information processing \cite{Witthaut2012EL}. The inherent potential of the system is illustrated in Fig. \ref{g2_fig}(d), which shows the expected amount of photon bunching after deconvoluting the slow decoherence processes found in the experiment. For the parameters of the present experiment and in the weak excitation limit, we predict $g^2(0) \sim 2.1$, which is mainly limited by the pure dephasing rate $\gamma_0.$ Importantly, such decoherence has been shown to give a minor contribution in single-photon indistinguishability measurements on quantum dots controlled by electrical gates  \cite{Matthiesen2013NCOM}, i.e., even more dramatic photon-photon scattering processes should be obtainable.

Having access to single-photon nonlinearities may open new perspectives for processing both classical and quantum information encoded in photons. In the classical regime, it enables constructing ultimately energy-efficient optical switches that are triggered by just a few quanta of light, which is required to outperform electronic transistors \cite{miller2010optical}. In the quantum regime, it may enable new and resource-efficient functionalities  required for deterministic quantum-information processing with photons \cite{Ralph2015underpreparation}. With spectral and spatial control of the quantum dots \cite{Jamil2014APL} the present system can be scaled to obtain multiple quantum dots deterministically coupled to a photonic-crystal waveguide, each inducing a giant nonlinearity. Such a complex quantum system may be exploited for advanced quantum simulations and to engineer novel quantum states of coupled light and matter \cite{chang2008crystallization}.

We gratefully acknowledge financial support from the Villum Foundation, the Carlsberg Foundation, the Lundbeck foundation, the Danish Council for Independent Research (Natural Sciences and Technology and Production
Sciences), and the European Research Council (ERC Consolidator Grant ALLQUANTUM). JDS and EHL acknowledge the support from the GRL and KIST institutional programs.

% (*** Consider the option of plotting and discussing NL phase shift. Consider specifying switching energy (probably attoJ), check Mabuchi ***)

\bibliography{bigBib}
\end{document}